\documentclass{svproc}

\usepackage{xcolor}
\usepackage{amsmath,amssymb,amsfonts}
\usepackage{graphicx}
\usepackage{threeparttable}
\usepackage{float}
\usepackage{mathtools}
\usepackage[ampersand]{easylist}
\usepackage[nottoc]{tocbibind}
\usepackage{enumitem}

\graphicspath{ {./fig/} }

\newif\ifshowcomments
\showcommentstrue

\ifshowcomments
\newcommand{\mynote}[2]{\fbox{\bfseries\sffamily\scriptsize{#1}}
	{\small$\blacktriangleright$\textsf{\emph{#2}}$\blacktriangleleft$}}
\else
\newcommand{\mynote}[2]{}
\fi

\begin{document}
	
\mainmatter

\title{A Methodology to Select Topology Generators for WANET Simulations\\(Extended Version)}



\author{Michael O'Sullivan \and Leonardo Aniello \and Vladimiro Sassone}
\institute{Electronics and Computer Science, \textit{University of Southampton}, Southampton, UK\\
\email{\{M.O'Sullivan | L.Aniello | vsassone\}@soton.ac.uk}}


\maketitle


\begin{abstract}

Many academic and industrial research works on WANETs rely on simulations, at least in the first stages, to obtain preliminary results to be subsequently validated in real settings. Topology generators (TG) are commonly used to generate the initial placement of nodes in artificial WANET topologies, where those simulations take place. The significance of these experiments heavily depends on the representativeness of artificial topologies. Indeed, if they were not drawn fairly, obtained results would apply only to a subset of possible configurations, hence they would lack of the appropriate generality required to port them to the real world. Although using many TGs could mitigate this issue by generating topologies in several different ways, that would entail a significant additional effort. Hence, the problem arises of what TGs to choose, among a number of available generators, to maximise the representativeness of generated topologies and reduce the number of TGs to use.

In this paper, we address that problem by investigating the presence of bias in the initial placement of nodes in artificial WANET topologies produced by different TGs. We propose a methodology to assess such bias and introduce two metrics to quantify the diversity of the topologies generated by a TG with respect to all the available TGs, which can be used to select what TGs to use. We carry out experiments on three well-known TGs, namely BRITE, NPART and GT-ITM. Obtained results show that using the artificial networks produced by a single TG can introduce bias. 


\end{abstract}


\section{Introduction} \label{s:intro}

A \textit{wireless ad hoc network} (WANET) is based on a decentralised topology of devices/nodes that cooperate to implement some routing protocol, i.e. each device forwards its own and other devices' traffic according to a specific algorithm with the aim of reaching the target destination. WANETs do not rely on any fixed infrastructure and each node can only communicate with those other nodes lying within the transmission range of one another. WANET applications are wide and significant, ranging from wireless sensor networks to vehicular ad hoc networks (VANETs) to mobile ad hoc networks (MANETs), and they are used in everyday scenarios as well as more critical settings, such as military operations.

Several WANET aspects are still being investigated by the research community, e.g. routing protocols~\cite{7793459,8109405} and security~\cite{7510878,7925687}. For convenience, many academic works heavily rely on simulation to test a proposed solution and obtain preliminary results that are used to validate its effectiveness. Network simulations commonly entail evaluating a given approach on many different WANET topologies to ensure results are meaningful, i.e. to have evidence that they can apply to a wide variety of networks and are not tied to particular network configurations. Hence, as also suggested by G\"unes et al.~\cite{gunes2011}, a key aspect in any network protocol simulation is the design and selection of what test network topologies to consider.

Network \textit{topology generators} (TGs) are usually employed to create a possibly large number of topologies, on the basis of predefined network models, real-world measurements and additional parameters available to tune the generation process. Although any TG is designed and implemented to generate a representative set of topologies, different TGs do not rely on the same models and assumptions, do not follow the same generation approach and thus are likely to produce diverse topologies, which in turn can lead to obtain dissimilar simulation results~\cite{Magonia,Heckmann2003}. Hence, we claim that the choice of the TG can affect this type of experiments, i.e. a TG is likely to introduce \textit{bias} in simulations. This holds true for WANET simulations as well, where TGs are used to generate the initial placement of nodes, which in turn plays an important role in the way a WANET network evolves over time.

Despite the fact that each TG has its own peculiarities, and that sometimes researchers can select a TG on the basis of the specific mathematical or physical model they need, there are in general several TGs that can be used to create artificial topologies representing the initial placement of nodes in WANETs. In this context, the best option would be to use all the available TGs to run simulations on the largest possible range of topologies, so as to ensure that obtained results are not biased by the choice of a specific TG, or subset of TGs. On the other hand, using many TGs proves to be really demanding for researchers in terms of required time and effort to delve into the technical issues of each TG. Therefore, a trade-off arises between reducing the effort to spend in setting up the simulations, i.e. minimising how many TGs to use, and maximising the representativeness of the simulations themselves, i.e. \textit{minimising the bias introduced by TG selection}.

In this paper, we delve into the analysis of the differences between topologies generated by distinct TGs to help researchers to reduce how many TGs to use while still preserving the representativeness of generated topologies. In particular, \textit{given a fixed number of available TGs}, we address the following research questions.
\begin{itemize}
	\item \textbf{RQ1}: How to measure the difference between topologies generated by distinct TGs? i.e. how to characterise the bias introduced by the choice of a specific TG rather than using all the TGs?
	\item \textbf{RQ2}: how to choose what TG, or TGs, to use to reduce such a bias?
\end{itemize}

\medskip

The approach we propose relies on a compact, numeric representation of topologies, based on a number of aspects about how network nodes are placed over the plane (e.g. inter-node distance, clustering) and about how WANETs work (e.g. nodes can only communicate with other nodes within their transmission range). Each topology is modelled as a vector of numeric features, which enables to compute distance metrics. We consider a fixed number of TGs and propose to interpret the bias as a measure of the differences that arise in generated WANET topologies when selecting any single TG, or subset of TGS, instead of picking all the available TGs.

We tackle RQ1 by focussing on two complementary facets of the distances between topologies. On the one hand, we want to \textit{quantify} the bias by measuring the average distance between topologies generated by distinct TGs. In the specific, we use \textit{Hedges' g} measure of effect size to compute the \textit{bias index}, which measures the \textit{difference between topologies produced by a specific TGs, or subset of TGs, and those created by all the available TGs}. On the other hand, we are also interested in evaluating to what extent existing differences are distinguishing of some TG, i.e. whether such differences allow to determine which TG generated a topology, regardless of the extent of those differences. In this regard, we employ machine learning techniques to compute the \textit{classification accuracy}, i.e. to estimate how precisely we can discover which TG generated a topology. 

We answer RQ2 by proposing a simple methodology, based on the bias index, to select what TGs to use to reduce the bias, depending on how many TGs can be picked at most.

We carry out an experimental evaluation using three well-known TGs, i.e. BRITE, NPART and TG-ITM. Obtained results show that using a single TG is likely to introduce bias, and that in this case picking NPART is the best choice to mitigate this issue. If two TGs can be used, BRITE and NPART provide the lowest bias. The experiments on the classification accuracy show that topologies can be correctly classified according to their TGs with high accuracy, i.e. up to almost 78\%, and that, in this specific case,  four topology features contribute most to distinguishing between different TGs.

To the best of our knowledge, this is the first work in literature that systematically investigates the differences between topologies generated by diverse TGs in the context of WANET simulation. The contributions of this work are
\begin{enumerate}
	\item the definition of a \textit{vector-based representation of WANET topologies}, based on a number of features derived from different aspects of node placement;
	\item the definition of two novel metrics to assess the differences between TGs, i.e. the \textit{bias index} and the \textit{classification accuracy};
	\item a methodology to choose what TG, or TGs, to use among available TGs to minimise the bias;
	\item an experimental evaluation on BRITE, NPART and GT-ITM TGs, showing the presence of bias in picking either a single TG or a pair of TGs.
\end{enumerate}

\medskip

The rest of the paper is organised as follows. Section~\ref{s:rel_work} describes background and discusses related work. The system model for our investigation is introduced in section~\ref{s:system_model}. The methodology we propose is detailed in section~\ref{s:methodology}. The experiments and obtained results are presented in section~\ref{s:evaluation}. Finally, section~\ref{s:conclusion} draws conclusions and outlines possible future work.

\section{Background and Related Work} \label{s:rel_work}

In this paper we focus on TGs that provide the initial placement of nodes over a plane. As we are dealing with WANETs, we are not interested in how nodes are connected among each other and assume that any node can communicate directly with all the nodes lying within its transmission range.

TGs can differ mainly in how nodes placement is decided~\cite{Sanni2013} and what each node represents~\cite{Heckmann2003}.
\textit{Node placement strategy} can be based either on some \textit{predefined model} or on \textit{real-world measurements}. In the former case, a certain probability distribution can be used, such as the \textit{Waxman model}~\cite{Waxman1988}, or specific strategies can be enforced to preserve the inter-node distance among nodes placed on a line (\textit{chain node placement}) or to position nodes at the intersections of square cells when the plane is organised as a grid (\textit{grid node placement}). In the latter case, nodes positions are instead determined in compliance with real-world measurements of existing network topologies.
%
%
Nodes in an artificial topology can represent either autonomous systems (AS), i.e. \textit{AS-level topologies}, or routers, i.e. \textit{router-level topologies}.

\medskip

Some existing works in literature deal with the investigation of diverse aspects of TGs, e.g. how realistic generated topologies are.
%
%
%
%
Several works~\cite{Magoni,Magoni2006,Magonia} focus on TGs for Internet topologies by comparing the topologies they generate with available real Internet map topologies, with the aim of assessing to what extent those topologies can be considered realistic.
Rossi et al.~\cite{Rossi2013} propose a framework to analyse Internet topologies by using a multi-level approach based on a number of graph measures and existing reference datasets. Their goal is to assess whether Internet TGs comply with their claimed objectives and how realistic generated topologies are.
Our work differs from those papers mainly because we do not evaluate whether artificial topologies are realistic, rather we investigate the bias in topologies generated by different TGs. Furthermore, we tackle WANETs rather than Internet. 

Heckmann et al.~\cite{Heckmann2003} compare three TGs according to the similarity of generated topologies with an available collection of real-world topologies.


Although all those works, likewise ours, focus on evaluating and comparing existing TGs, the main difference lies in the goal of such a comparison. In fact, while existing literature is interested in measuring how well generated topologies represent real-world networks, we concentrate on an orthogonal aspect by investigating whether picking a certain TG rather than another one, or rather than choosing more TGs, can introduce bias. 
From this point of view, our contribution is novel and complements existing research on comparing available TGs.

\section{System Model} \label{s:system_model}

We consider a set $\mathcal{TG}$ with $N^{TG}$ topology generators (TG), i.e. $|\mathcal{TG}| = N^{TG}$. Each TG generates coordinates for the initial placement of nodes, i.e. devices, within a defined square \textit{topology area}, with sides $D$ \textit{units} long. 
%
%
Each TG $tg_i$ generates a set $\mathcal{T}_i$ with $N^T$ topologies, where $i{=}0, \dots, N^{TG}{-}1$.
The set containing all the topologies generated by all the TGs is referred to as 
\begin{equation*}
\mathcal{T} = \bigcup_{i=0}^{N^{TG}{-}1}\mathcal{T}_i
\end{equation*}
hence $|\mathcal{T}| = N^{TG} \cdot N^T$.
%
%
Each topology $t_j \in \mathcal{T}_i$ has $N$ nodes $\mathcal{N}_j = \{n_k\}$, where $i{=}0, \dots, N^{TG}{-}1$, $j{=}0, \dots, N^T{-}1$, $k{=}0, \dots, N{-}1$.
Each node $n_k$ is identified by its bi-dimensional coordinates $(x_k, y_k)$ in the topology area, where $0 \le x_k,y_k \le D$.
Given two nodes $n_a$ and $n_b$ ($a,b{=}0, \dots, N{-}1$), we define their Euclidean distance as
\begin{equation*}
d(n_a,n_b) = \sqrt{(x_a-x_b)^2 + (y_a-y_b)^2}
\end{equation*}

In WANETs, any device can establish connections with other devices placed within a specific distance, which we refer to as \textit{radius} $r$. We consider a number $N^R$ of different radii $\mathcal{R} = \{r_i\}$, where $i{=}0, \dots, N^R{-}1$ and $0<r_j<r_{j+1}<D$ for $j{=}0, \dots, N^R{-}2$. 



\section{Methodology for Analysing the Bias of Topology Generators} \label{s:methodology}

In general, a topology generator (TG) introduces bias if the topologies it generates are not representative enough of some target application, such as analysing routing protocols. It is not trivial to decide whether a given set of topologies can be considered representative enough of a certain application, let alone it is possible to provide general criteria to evaluate the representativeness of a group of topologies regardless of what they are intended to be used for. However, if we consider the universe set $\mathcal{TU}$, containing all the possible topologies, and a subset of it $\mathcal{S} \subset \mathcal{TU}$, we can investigate to what extent $\mathcal{S}$ is representative of $\mathcal{TU}$ by inspecting the differences between topologies in $\mathcal{S}$ and topologies in $\mathcal{TU}$. We propose to use those differences to analyse the bias of using topologies in $\mathcal{S}$ only, i.e. the \textit{larger} and \textit{sharper} such differences, the higher the bias.

Although we cannot have in practice a set like $\mathcal{TU}$, we do have a number of available TGs, $\mathcal{TG}$ (see section~\ref{s:system_model}), which can be used to generate a set of topologies $\mathcal{T}$. While we do not know how much $\mathcal{T}$ is representative of $\mathcal{TU}$, we claim that $\mathcal{T}$ is the best approximation of $\mathcal{TU}$ we can aim for from a pragmatic point of view. Hence, to measure the bias introduced by a TG $tg_i \in \mathcal{TG}$, we can examine the differences between the topologies it generates, $\mathcal{T}_i$, and the topologies in $\mathcal{T}$.

\medskip

We propose a two-steps methodology to analyse the bias of TGs. 
The first step is modelling topologies by extracting a number of characteristic features, which will be used to have a compact, numeric representation of topologies and enable to measure the differences between them.
The second step is indeed computing metrics to quantify the dissimilarities between topologies generated by different TGs.
We claim that there are two complementary aspects to investigate when analysing such dissimilarities. On the one hand, the extent of those differences, i.e. how large they are, provides an objective scale of the bias. On the other hand, the peculiarity of those differences, i.e. how much distinctive of TGs they are, allows to figure out whether topologies generated by different TGs are distinguishable from each other by looking at specific aspects, regardless of the extent of the existing differences.
We propose the following two approaches.
\begin{itemize}
	\item Computing the average distance between the topologies generated by a TG and all the topologies in $\mathcal{T}$. By mapping topologies into the space generated by the chosen features, we use the Hedges' g~\cite{Hedges1981}, measure of effect size, to quantify the difference between two populations: the topologies generated by a specific TG and the topologies in $\mathcal{T}$. We refer to such a difference as \textit{bias index}.
	\item Assessing the accuracy in distinguishing which TG generated a given topology. We train a classifier with the topologies in $\mathcal{T}$ and the information on which TG generated each of them, then we test the obtained classification model by measuring its \textit{classification accuracy}.
\end{itemize}

We define the features we use to characterise topologies in section~\ref{s:topology_features}, then we detail how compute the bias index and the classification accuracy in section~\ref{s:bias_index} and~\ref{s:classification_accuracy}, respectively,

\subsection{Topology features} \label{s:topology_features}

To choose what features to consider, we focus on the aspects we deem most representative to show variance within topologies. We thus consider the features that characterise the placement of the nodes within the test plane and relationships between nodes. We extract features by looking at the following aspects of a generic topology $t_j$: inter-node distance, node spatial distribution, node density, shared node neighbours, node clustering coefficient.

\medskip \noindent \textbf{Inter-node Distance.}
We consider the set of node distances $\mathcal{D}$ defined as follows
\begin{equation*}
	\mathcal{D} = \{d(n_a,n_b) \mid n_a,n_b \in \mathcal{N}_j, 0 \le a < b < N\}
\end{equation*}
The features we extract from $\mathcal{D}$ are (i) the minimum value $d_{min}$, (ii) the maximum value $d_{max}$, (iii) the value range $d_{max} - d_{min}$, (iv) the mode, i.e. the most frequent value~\footnote{If there are more most-frequent values, by convention we pick the smallest one.}, (v) how many times the mode occurs, (vi) the mean value and (vii) the standard deviation.

\begin{figure}[h]
	\centering
	\includegraphics[width=0.4\columnwidth]{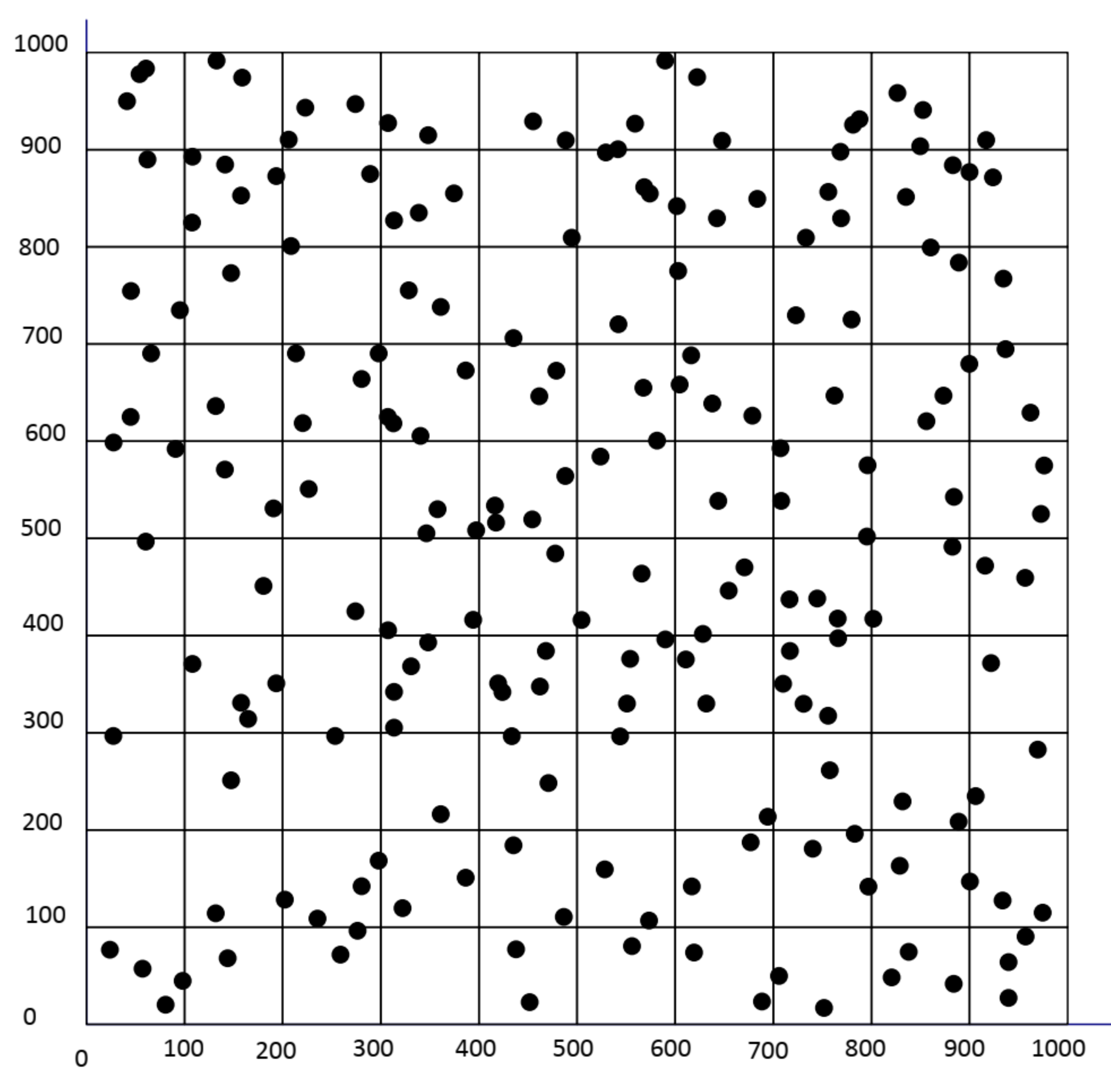}
	\caption{Partition of a topology area with 1000 units sides in 100 smaller squares, each with 100 units dies. This partitioning is used to compute spatial distribution features.}
	\label{f:spatial_distribution}
\end{figure}

\medskip \noindent \textbf{Spatial Distribution.}
By taking inspiration from the Quadrat Method~\cite{greig1952use} used to test the Complete Spatial Randomness hypothesis, we partition the topology area in $d^2$ smaller squares, each with sides $D/d$ units long.
Figure~\ref{f:spatial_distribution} shows an example of topology area partitioning.

We consider the set $\mathcal{NC}$ of node counts, where each element $nc_s \in \mathcal{NC}$ is the number of nodes in the $s$-th partition of the topology area, with $s{=}0, \dots, d^2{-}1$.
The features we extract from $\mathcal{NC}$ are (i) the minimum value $nc_{min}$, (ii) the maximum value $nc_{max}$, (iii) the value range $nc_{max} - nc_{min}$, (iv) the mode and (v) how many times the mode occurs.

\medskip \noindent \textbf{Node Density.}
We define the density $nd_r(n_a)$ of a node $n_a \in \mathcal{N}_j$ of a topology $t_j$, for a given radius $r \in \mathcal{R}$, as the number of other nodes within distance $r$ from $n_a$, i.e.
\begin{equation*}
	nd_r(n_a) = |\{n_b \in \mathcal{N}_j \setminus \{n_a\} \mid d(n_a,n_b) < r \}|
\end{equation*}
We extract as many features $f_{density}$ as the number of radii in $\mathcal{R}$, each corresponding to the average node density for a given radius $r$, defined as follows
\begin{equation*}
	f_{density}(r) = \frac{\sum_{n_a \in \mathcal{N}_j}nd_r(n_a)}{N}, r \in \mathcal{R}
\end{equation*}

\begin{figure}[h]
	\centering
	\includegraphics[width=0.4\columnwidth]{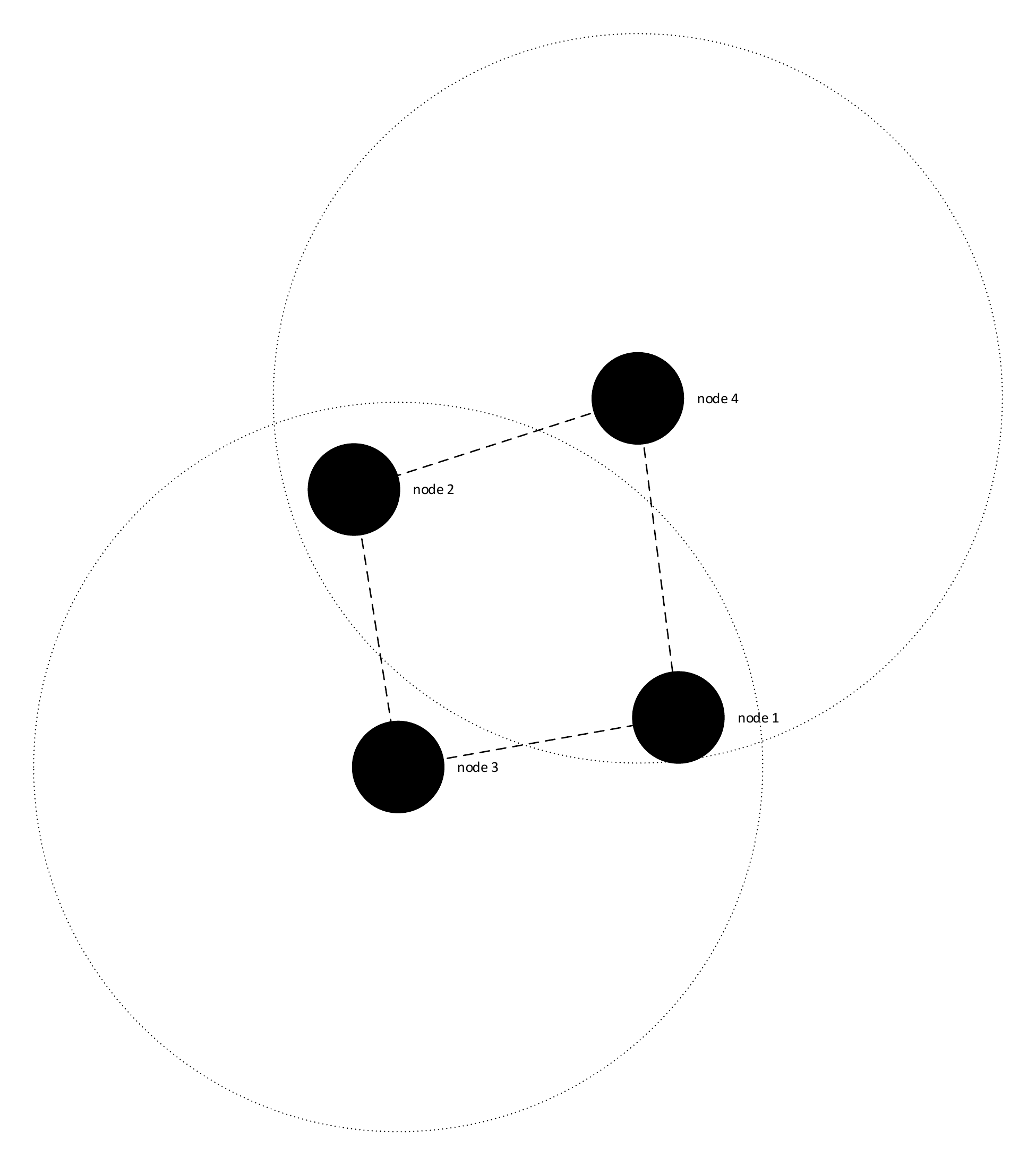}
	\caption{Nodes 1 and 2 are shared neighbours of nodes 3 and 4. In this case, the value of \textit{shared neighbours count} for nodes 3 and 4 is 2.}
	\label{f:shared_neighbours_distribution}
\end{figure}

\medskip \noindent \textbf{Shared Neighbours Distribution.}
For any given pair of nodes $n_a,n_b \in \mathcal{N}_j$ and radius $r \in \mathcal{R}$, the \textit{shared neighbours}~\cite{Nowak2014} are those nodes within distance $r$ from both $n_a$ and $n_b$. Figure~\ref{f:shared_neighbours_distribution} shows an example where nodes 1 and 3 are shared neighbours of nodes 3 and 4.
We first introduce the neighbours function $neigh_r(n_a)$ for a node $n_a \in \mathcal{N}_j$ and a radius $r$ as
\begin{equation*}
	neigh_r(n_a) = \{n_b \in \mathcal{N}_j \setminus \{n_a\} \mid d(n_a,n_b) < r \}
\end{equation*}
Then we define the shared neighbours count $snc_r(n_a,n_b)$ for nodes $n_a,n_b \in \mathcal{N}_j$ and radius $r \in \mathcal{R}$ as
\begin{equation*}
	snc_r(n_a,n_b) = |neigh_r(n_a) \cap neigh_r(n_b)|
\end{equation*}
We extract as many features $f_{shared\_neigh}$ as the number of radii in $\mathcal{R}$, each corresponding to the average shared neighbours count for a given radius $r$, defined as follows
\begin{equation*}
f_{shared\_neigh}(r) = \frac{\sum_{n_a,n_b \in \mathcal{N}_j, n_a \neq n_b}snc_r(n_a,n_b)}{N(N-1)/2}, r \in \mathcal{R}
\end{equation*}

\begin{figure}[h]
	\centering
	\includegraphics[width=0.4\columnwidth]{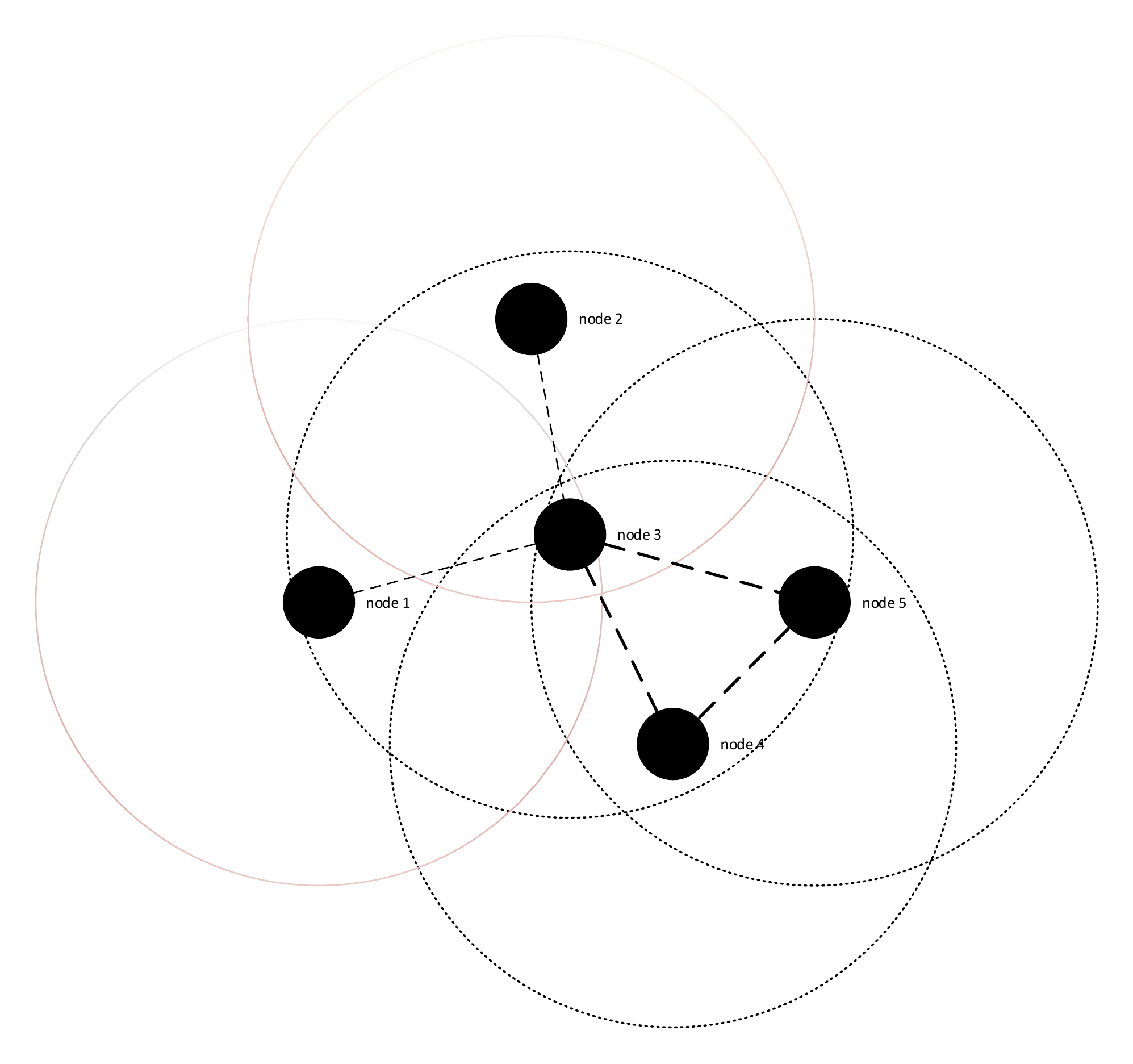}
	\caption{The neighbours of node 3 are nodes 1, 2, 4 and 5. Among those neighbours, there is a pair of nodes, i.e. nodes 4 and 5, which are neighbours of each other, while nodes 1 and 2 are not neighbour of any other node.}
	\label{f:clustering_coefficient}
\end{figure}

\medskip \noindent \textbf{Clustering coefficient.}
The clustering coefficient~\cite{Holland1971} of a node $n_a \in \mathcal{N}_j$ is a measure based on the number $c$ of node pairs that lie within distance $r \in \mathcal{R}$ from $n_a$ and are neighbours of each other. An example is reported in figure~\ref{f:clustering_coefficient}.
This coefficient is calculated as the ratio between $c$ and the number of neighbours of $n_a$, i.e. its density. More formally the clustering coefficient $cc_r(n_a)$ of a node $n_a \in \mathcal{N}_j$ for a radius $r \in \mathcal{R}$ is defined as
\begin{equation*}
	cc_r(n_a) = \frac{|\{n_b,n_c \in neigh_r(n_a) | 0 < d(n_b,n_c) < r\}|}{nd_r(n_a)}
\end{equation*}
We extract as many features $f_{clustering}$ as the number of radii in $\mathcal{R}$, each corresponding to the average cluster coefficient for a given radius $r$, defined as follows
\begin{equation*}
f_{clustering}(r) = \frac{\sum_{n_a \in \mathcal{N}_j}cc_r(n_a)}{N}, r \in \mathcal{R}
\end{equation*}

\subsection{Bias Index} \label{s:bias_index}

The bias index of a TG $tg_i \in \mathcal{TG}$ with respect to all the TGs in $\mathcal{TG}$ is measured as the distance between the topologies $\mathcal{T}_i$ generated by $tg_i$ and all the topologies generated $\mathcal{T}$. This distance is computed on the basis of the following feature-based representation of a topology $t_j$
\begin{equation*}
	t_j = \langle f^j_0, \dots, f^j_{F-1} \rangle
\end{equation*}
where $f^j_k$ is the value of the $k$-th feature of $t_j$ ($k=0, \dots, F{-}1$) and $F$ is the number of used features, detailed in section~\ref{s:topology_features}, equal~\footnote{There are 7 features for inter-node distances, 5 features for spatial distribution and as many features as the number of radii $N^R$ for (i) node density, (ii) shared neighbours distribution and (iii) clustering coefficient (see section~\ref{s:topology_features}).} to $12 + 3N^R$.

\textit{Hedges’ g}~\cite{Hedges1981} is used to estimate the standardised mean difference between two populations, i.e. the average distance between the elements of two different populations, measured in standard deviations. Although in its original form it can be applied to single-dimension elements only, we propose to extend Hedges' g to $F$ dimensions to quantify the difference between topologies in $\mathcal{T}_i$ and in $\mathcal{T}$.

We first detail how to apply Hedges' g to a single feature $f_k$, where $k=0, \dots, F{-}1$. We define $\mathcal{T}^k$ and $\mathcal{T}_i^k$ as the projections of $\mathcal{T}$ and $\mathcal{T}_i$ to feature $f_k$, respectively, as follows
\begin{equation*}
	\mathcal{T}^k = \{f^j_k \mid t_j = \langle f^j_0, \dots, f^j_{F-1} \rangle \in \mathcal{T} \}
\end{equation*}
\begin{equation*}
	\mathcal{T}_i^k = \{f^j_k \mid t_j = \langle f^j_0, \dots, f^j_{F-1} \rangle \in \mathcal{T}_i \}
\end{equation*}
Let $m^k$ and $s^k$ be the mean and standard deviation of $\mathcal{T}^k$, respectively.
Let $m_i^k$ and $s_i^k$ be the mean and standard deviation of $\mathcal{T}_i^k$, respectively.
In compliance with the original formulation, we define Hedges' g for a single feature $f_k$ as
\begin{equation*}
	g_i^k = \frac{m^k - m_i^k}{s_i^{*k}}
\end{equation*}
where $s_i^{*k}$ is the pooled standard deviation for $\mathcal{T}^k$ and $\mathcal{T}_i^k$, computed as follows
\begin{equation*}
	s_i^{*k} = \sqrt{ \frac{(|T^k| - 1) \cdot (s^k)^2 + (|T_i^k| - 1) \cdot (s_i^k)^2}{|T^k| + |T_i^k| - 2}}
\end{equation*}
Finally, to obtain the bias index $g_i$ for $tg_i$, we combine all the $F$ values $g_i^k$ by considering each of them as a distance along one dimension, as follows
\begin{equation*}
	g_i = \sqrt{\sum_{k=0}^{F-1}(g_i^k)^2}
\end{equation*}

\medskip

\noindent \textbf{TGs Selection.}
The bias index can be used to choose what TG to pick to reduce the possible bias. Selecting the TG with the lowest bias index would correspond to using the set of topologies with the lowest distance, on average, from the whole set $\mathcal{T}$ of available topologies. According to the methodology approach introduced at the beginning of this section, this in turn means choosing the most representative subset of topologies available, if a single TG has to be selected.

If more than one TG can be picked, say $p$ out of $N^{TG}$, then the same strategy can be used by considering the possible $\binom{N^{TG}}{p}$ subsets of $\mathcal{T}$, each in the form
\begin{equation*}
	\mathcal{T}_{i_0, \dots, i_{p-1}} = \bigcup_{j=0}^{p-1} \mathcal{T}_{i_j}
\end{equation*}
with $\mathcal{T}_{i_j} \subset \mathcal{T}$, $0 \le i_0 < \dots < i_{p-1} < N^{TG}$, $0 < p < N^{TG}$, and computing the corresponding bias index. Again, the subset with the lowest bias index is the most representative of $\mathcal{T}$. We refer to $g_{i_0, \dots, i_{p-1}}$ as the bias index of $\mathcal{T}_{i_0, \dots, i_{p-1}}$.

\subsection{Classification Accuracy} \label{s:classification_accuracy}

We consider the accuracy of a classifier trained with the \textit{topology generator ground truth} $\mathcal{TGGT}$ defined as follows
\begin{equation*}
	\mathcal{TGGT} = \bigcup_{i=0}^{N^{TG}{-}1} \{\langle t_j, tg_i \rangle \mid t_j \in \mathcal{T}_i \}
\end{equation*}
where each pair $\langle t_j, tg_i \rangle$ represents the fact that topology $t_j$ has been generated by TG $tg_i$, i.e. $tg_i$ is the \textit{class} of $t_j$.
We use part of the ground truth for training and the other for testing, where we compute the actual classification accuracy.
To avoid any possible bias deriving from the choice of how the ground truth is split between training and testing, we employ the well-known $k$-fold cross validation~\cite{stone1974cross} method, which works as follows.
The ground truth is first partitioned in $k$ equally sized folds, then a classifier is trained from the scratch in $k$ different ways by using each time all the folds but one. After each training, the resulting classifier is tested by using the fold excluded during the training and the classification accuracy is recorded. The final accuracy is the average of the $k$ accuracy values obtained during the $k$ trainings.

More formally, let $\mathcal{TGGT}_l$ be the $l$-th partition of $\mathcal{TGGT}$, for $l=0, \dots, k{-}1$, i.e.
\begin{equation*}
	\mathcal{TGGT} = \bigcup_{l=0}^{k-1} \mathcal{TGGT}_l
\end{equation*}
\begin{equation*}
\mathcal{TGGT}_a \cap \mathcal{TGGT}_b = \emptyset , \quad 0 \le a < b < k
\end{equation*}
Since all the folds have the same size, we have that $|\mathcal{TGGT}_l| = N^T N^{TG} / k$.
Let $\mathcal{C}_l \colon \mathcal{T} \to \mathcal{TG}$ be the function computed by a classifier trained using all the folds except $\mathcal{TGGT}_l$. We define the classification accuracy $a_l$ for the $l$-th fold as the ratio of correctly classified topologies to the total number of classified topologies
\begin{equation*}
	a_l = \frac{|\{ \langle t_j, tg_i \rangle \in \mathcal{TGGT}_l \mid \mathcal{C}_l(t_j) = tg_i \}|}{|\mathcal{TGGT}_l|}
\end{equation*}
The final classification accuracy is defined as $a = \frac{\sum_{i=0}^{k-1}a_l}{k}$.
In the absence of bias, the classification accuracy should be close to $1/T^{TG}$. Higher values indicate that there are some features that are peculiar to specific TGs. In that case, a feature analysis can help to identify those features, to understand whether and to what extent they are relevant for the particular experiments the generated topologies have to be used for.
The feature analysis can be based on a \textit{sequential feature selector}~\cite{Aha1996} algorithm, used to reduce the initial dimension of the feature space. The goal is to create a subset of features that explain the most variance in the dataset. This is done by either adding or removing one feature at a time and measuring the corresponding classification accuracy until convergence is achieved, i.e. the process stops when the accuracy ceases to grow.

\section{Experimental Evaluation} \label{s:evaluation}

We apply the proposed methodology to a number of well-known TGs, detailed in section~\ref{s:topology_generators}. The parameters we choose to instantiate the model (see section~\ref{s:system_model}) are reported in section~\ref{s:evaluation_settings}. The experiments on bias index and classification accuracy, as well as obtained results, are described in sections~\ref{s:bias_index_evaluation} and~\ref{s:classification_accuracy_evaluation}, respectively. We also carry out a more detailed analysis on the impact of each feature on experiments outcomes, pointed out in section~\ref{s:feature_analysis}. 

\subsection{Topology Generators} \label{s:topology_generators}
In our experiments, we use three well known TGs: BRITE, NPART and GT-ITM, described in this section.

\medskip \noindent \textbf{BRITE.}
The \textit{Boston University Representative Internet Topology}~\cite{Medina2001} (BRITE) is a universal model-based TG designed to be extendable to enable the addition of new models. BRITE uses various models for the placement of nodes, as detailed below. 
\begin{enumerate}
	\item \textit{Flat Router model}. It represents a router-level topology and is designed for router networks.
	The placement of nodes is either random or uses the \textit{heavy tailed} approach, where the plane is divided into squares, each square is assigned a number of nodes drawn from a heavy-tailed distribution, and then nodes are placed randomly within the square.
	\item \textit{Flat AS-level model}. It is very similar to the Flat Router model except that it generates AS-level topologies. 
	\item \textit{Hierarchical Topologies model}. It generates Internet-like topologies. It can be configured to use either a \textit{top down} or \textit{bottom up} approach. In the first case, an AS-level topology is first built by using the Flat AS-level model, then for each node a router level topology is generated. In the second case, a router-level topology is first generated, then AS nodes are introduced and each is linked to a number of router nodes.
\end{enumerate}
The Flat AS-Level model is used for our experiments because it is more representative of a mesh network. 

\medskip \noindent \textbf{NPART.}
The \textit{Node Placement Algorithm for Realistic Topologies}~\cite{Milic2009} (NPART) TG generates topologies based on properties of real networks, i.e. any artificial network is generated randomly but in compliance with a number of properties of the real-world topologies. The generation algorithm used by NPART relies on a number of sociological and technological observations introduced by Aha et al.~\cite{Aha1996}, listed below.
\begin{itemize}
	\item It is more likely that a new participant joins the network in areas where connectivity is high.
	\item A participant in the network expects to have at least one single communicating link to the rest of the network, possibly creating a large number of penned nodes.
	\item A pendant node may become a seed for a new, larger and well-connected sub network.
	\item It is the network that specifies the area it occupies, not the other way around. So, instead of defining the node placement area like most of the existing placement algorithms, the network should be allowed to grow.
\end{itemize}
Configuration options for the generation of topologies are \textit{NPART Berlin} (based on the real Berlin's mesh network consisting of 275 nodes), \textit{NPART Leipzig} (based on the real Leipzig's mesh network consisting of 346 nodes), \textit{Uniform placement model} (it uses uniform probability placement within the test area), \textit{Grid placement} (also known as mesh placement, where nodes are located at intersection of a rectangular grid), \textit{Quasi-Grid placement} (node are placed as a Gaussian distribution with the mean given by regular grid points) and \textit{Random Waypoint Model} (it is a random model for the movement of mobile users and how their location, velocity and acceleration changes over time).

The uniform placement model has been used in our experiments.

\medskip \noindent \textbf{GT-ITM.}
The \textit{Georgia Tech Internetwork Topology Model}~\cite{calvert1997modeling} (GT-ITM) is a model-based TG that produces wide area networks like topologies. Two models are available to decide the node placement: \textit{Flat Random Graphs} and \textit{Hierarchical}.

The Flat Random Graphs model distributes nodes randomly over the test plane. This model does not aim to reflect the real world, it is rather for simplicity. A variations of this model uses Waxman probability~\cite{Waxman1988} produce more realistic topologies. 

The Hierarchical model creates a topology by connecting smaller components together according to a larger scale structure. This suggests that this model has a propensity towards clustering, hence we choose the Flat Random Graph model which represents better WANETs.

\subsection{Evaluation Settings} \label{s:evaluation_settings}

With reference to the system model defined in section~\ref{s:system_model}, we consider the $N^{TG} = 3$ TGs described in the previous section, i.e. $\mathcal{TG} = \{$BRITE, NPART, GT-ITM$\}$, and generate $N^T = 1000$ topologies for each TG. Each topology has $N=1000$ nodes.
The reference topology area has sides $D = 1000$ units long.
We evaluate the following $N^R = 8$ radii: $\mathcal{R} = \{5, 10, 20, 30, 40, 60, 80, 100\}$.

\subsection{Bias Index Evaluation} \label{s:bias_index_evaluation}

We compute the bias index $g_i$ for each TG and $g_{i,j}$ for each pair of TGs, as described in section~\ref{s:bias_index}. The results are reported in table~\ref{t:bias_index}.
As can be noted, BRITE topologies seem to be significantly different from those generated by NPART and GT-ITM, and vice-versa, which suggests that using either TG alone would provide a set of topologies significantly different from the set including all the topologies. However, if only one TG has to be selected, NPART proves to generate topologies that are less different on average from those generated by all available TGs. If two TGs can be chosen, BRITE and NPART show to be the best pair to consider.


\begin{table}[h]
	\begin{center}
		\caption{Bias index of the considered TGs.}
		\label{t:bias_index}
		\begin{tabular}{|l|r|} 
			\hline
			\textbf{Topology Generator(s)} & \textbf{Bias index}\\
			\hline
			NPART & 1.890\\
			GT-ITM & 2.145\\
			BRITE & 4.282\\
			\hline
			BRITE + NPART & 0.908\\
			BRITE + GT-ITM & 0.976\\
			NPART + GT-ITM & 2.430\\
			\hline
		\end{tabular}
	\end{center}
\end{table}

\subsection{Classification Accuracy Evaluation} \label{s:classification_accuracy_evaluation}

In this context, classification accuracy is investigated to understand to what extent topologies can be distinguished with respect to their TG. The higher the classification accuracy, the sharper the differences between topologies generated by different TGs. Although classifiers are commonly selected and tuned to maximise classification accuracy, in this case we are only interested in verifying whether the accuracy can be relevantly higher than $1/3$, i.e. $1/N^{TG}$ (see section~\ref{s:classification_accuracy}). We choose Naive Bayes as classifier because of its simplicity and we test three different probability distributions, i.e. Gaussian, Bernoulli and Multinomial, to assess whether results are consistent regardless of the particular distribution. Table~\ref{t:classification_accuracy} shows the classification accuracy for the three algorithms, where it can be observed that all the values are significantly larger than $1/3$.

\begin{table}[h]
	\begin{center}
		\caption{Classification accuracy for the three classification algorithms.}
		\label{t:classification_accuracy}
		\begin{tabular}{|l|c|c|} 
			\hline
			\textbf{Algorithm}  &  \textbf{Classification Accuracy (\%)}\\
			\hline
			\hline
			GaussianNB & 77.95\\
			\hline
			BernoulliNB & 58.56\\ 
			\hline
			MultinomialNB & 70.07\\
			\hline
		\end{tabular}
	\end{center}
\end{table}


\subsection{Features Analysis} \label{s:feature_analysis}

\begin{figure}[h]
	\centering
	\includegraphics[width=0.7\columnwidth]{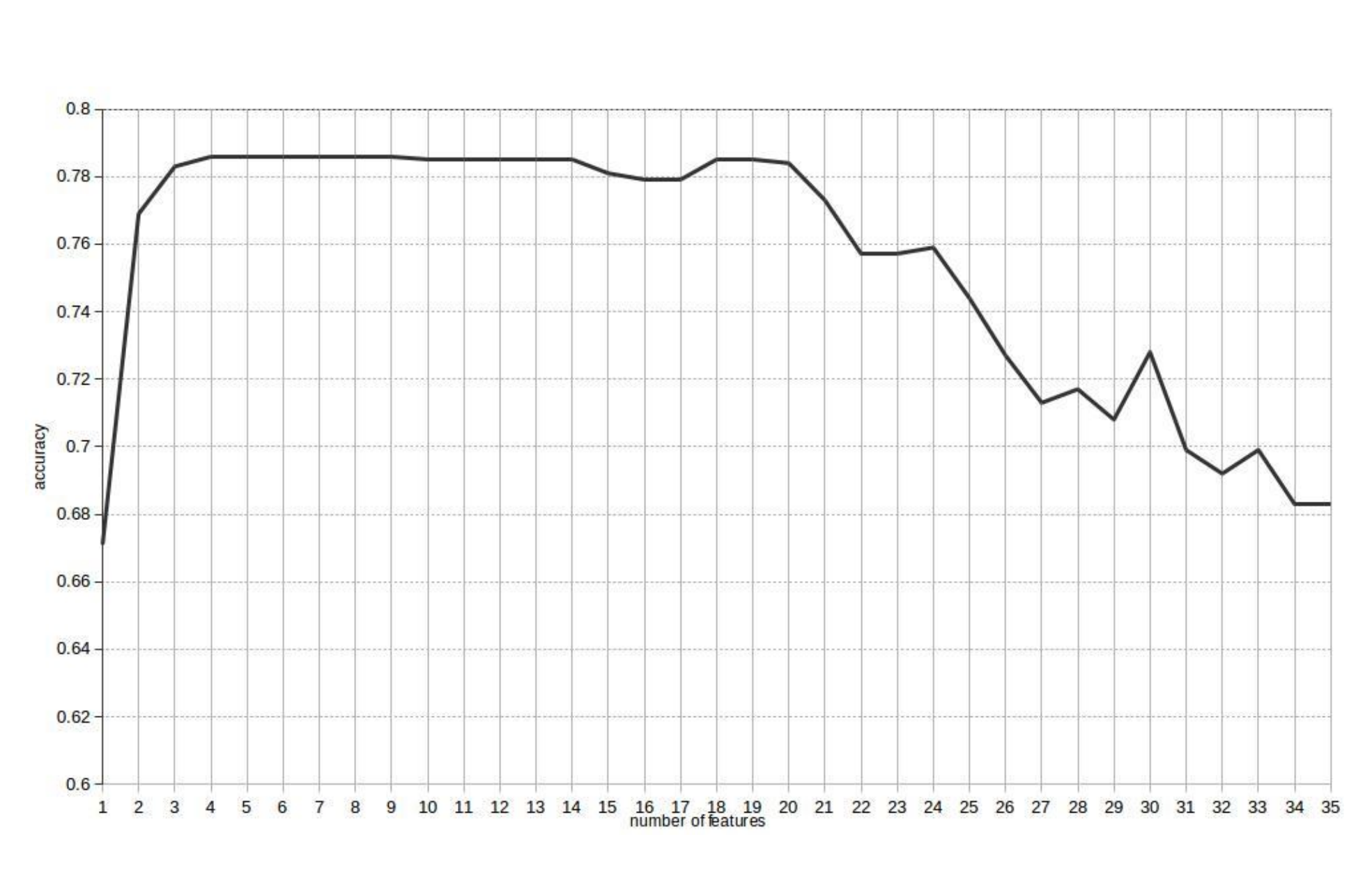}
	\caption{Classification accuracy by varying the number of used features for a specific fold.}
	\label{f:accuracy_varying_feature_count}
\end{figure}

We carry out a more detailed analysis of what features weight most for classification by applying the Forward Sequential Selection (FSS) method as sequential feature selector (see section~\ref{s:classification_accuracy}). FSS works sequentially, it starts with an empty set of features and, at each iteration, selects the feature that yields the highest accuracy. Figure~\ref{f:accuracy_varying_feature_count} shows how the classification accuracy varies with the number of considered features, for a single fold, and that the highest accuracy is achieved with 4 features.

Table~\ref{t:accuracy_varying_feature_count} details the outcomes of the first 9 iterations of FSS, in terms of classification accuracy and the corresponding set of features that yield such accuracy. Table~\ref{t:best_feature_set} lists the four features yielding the highest classification accuracy. This features analysis identifies what are the most distinguishing characteristics of topologies generated by different TGs, with respect to the three TGs used in our experiments. Whether this fact can introduce bias depends on the specific application those topologies have to be used for, i.e. to what extent those features are relevant for the target scenario.

\begin{table}[h]
	\caption{Classification accuracy and corresponding features set for the first 9 iterations of FSS.}
	\label{t:accuracy_varying_feature_count}
	\begin{center}
		\begin{tabular}{|c|c|l|} 
			\hline
			\textbf{Features} & \textbf{Accuracy} & \textbf{Feature IDs}\\
			\hline
			1 & 0.668 &31\\
			2 & 0.769 & 31,6\\
			3 & 0.783 & 31,6,30\\
			4 & \textbf{0.786} & \textbf{31,6,30,14}\\
			5 & 0.786 & 31,6,30,14,13 \\
			6 & 0.786 & 31,6,30,14,13,12\\
			7 & 0.786 & 31,6,30,14,13,12,15\\
			8 & 0.786 &  31,6,30,14,13,12,15,16\\
			9 & 0.786 & 31,6,30,14,13,12,15,16,20\\
			\hline
		\end{tabular}
	\end{center}
\end{table}

\begin{table}[h]
	\begin{center}
		\caption{Set of features yielding to the highest classification accuracy.}
		\label{t:best_feature_set}
		\begin{tabular}{|c|l|}  
			\hline
			\textbf{Feature ID}  & \textbf{Feature description}\\
			\hline
			31 & Shared neighbours distribution with 30 units radius\\
			6 &	Minimum inter-node distance\\
			30 & Shared neighbours distribution with 20 units radius\\
			14 & Node density with 20 units radius\\
			\hline
		\end{tabular}
	\end{center}
\end{table}

\section{Conclusion} \label{s:conclusion}

In this paper, we investigate the presence of bias in WANETs simulations due the choice of what TG, or TGs, to use among a fixed number of available TGs. We propose two metrics, namely bias index and classification accuracy, to measure the extent and significance of the distance between topologies generated by a single TG and topologies produced by all available TGs. We also propose a methodology to select what TG, or TGs, to pick to minimise the bias. 

We present an experimental evaluation where we compute bias index and classification accuracy for three well-known TGs: BRITE, NPART and GT-ITM. Obtained results prove that topologies generated by a single TG are different from those created by all the three TGs, and that the TG which generated a certain topology can be determined with relevant accuracy.

\medskip

As future work, we plan to carry out additional evaluations to investigate how bias index and classification accuracy are linked to variance in the results of same experiments performed on different TGs. A number of reference algorithms can be chosen, e.g. routing protocols, and executed on available generated topologies to verify whether lower values for bias index and classification accuracy can actually lead to reduced variance of obtained results, with respect to the experimental outcomes that would be achieved by using all the available TGs.

An additional, significant future work concerns the sensitivity analysis on both system model parameters and TGs configurations, to assess to what extent such a tuning affects computed values for bias index and classification accuracy. Furthermore, the choice of what classifier to use for the classification accuracy needs to be investigated by considering a larger number of classification algorithms.

\begingroup


\bibliographystyle{plain}

\bibliography{references}

\begin{thebibliography}{10}

\bibitem{Aha1996}
David~W Aha and Richard~L Bankert.
\newblock A comparative evaluation of sequential feature selection algorithms.
\newblock In {\em Learning from data}, pages 199--206. Springer, 1996.

\bibitem{calvert1997modeling}
Kenneth~L Calvert, Matthew~B Doar, and Ellen~W Zegura.
\newblock Modeling internet topology.
\newblock {\em IEEE Communications magazine}, 35(6):160--163, 1997.

\bibitem{7793459}
Bo~Cheng and G.~Hancke.
\newblock Energy efficient scalable video manycast in wireless ad-hoc networks.
\newblock In {\em IECON 2016 - 42nd Annual Conference of the IEEE Industrial
  Electronics Society}, pages 6216--6221, Oct 2016.

\bibitem{8109405}
C.~Cheng and S.~Lin.
\newblock A hole-bypassing routing algorithm for wanets.
\newblock In {\em 2017 IEEE 42nd Conference on Local Computer Networks (LCN)},
  pages 547--550, Oct 2017.

\bibitem{greig1952use}
Peter Greig-Smith.
\newblock The use of random and contiguous quadrats in the study of the
  structure of plant communities.
\newblock {\em Annals of Botany}, pages 293--316, 1952.

\bibitem{gunes2011}
M.~H. G\"unes and M.~B. Akgün.
\newblock Link-level network topology generation.
\newblock In {\em Proceedings of 31st International Conference on Distributed
  Computing Systems Workshops (ICDCSW)}, 2011.

\bibitem{Heckmann2003}
Oliver Heckmann, Michael Piringer, Jens Schmitt, and Ralf Steinmetz.
\newblock On realistic network topologies for simulation.
\newblock In {\em Proceedings of the ACM SIGCOMM workshop on Models, methods
  and tools for reproducible network research}, pages 28--32. ACM, 2003.

\bibitem{Hedges1981}
L.~V. Hedges.
\newblock {Distribution Theory for Glass's Estimator of Effect size and Related
  Estimators}.
\newblock {\em Journal of Educational and Behavioral Statistics}, 1981.

\bibitem{Holland1971}
Paul~W. Holland and Samuel Leinhardt.
\newblock {Transitivity in Structural Models of Small Groups}.
\newblock {\em Comparative Group Studies}, 1971.

\bibitem{Magoni}
D.~Magoni and J.~. Pansiot.
\newblock Evaluation of internet topology generators by power law and distance
  indicators.
\newblock In {\em Proceedings 10th IEEE International Conference on Networks
  (ICON 2002). Towards Network Superiority (Cat. No.02EX588)}, pages 401--406,
  2002.

\bibitem{Magoni2006}
Damien Magoni and J~J Pansiot.
\newblock {Analysis and Comparison of Internet Topology Generators}.
\newblock {\em NETWORKING 2002: Networking Technologies, Services, and
  Protocols; Performance of Computer and Communication Networks; Mobile and
  Wireless Communications}, 2345:364--375, 2006.

\bibitem{Magonia}
Damien Magoni and Jean-Jacques Pansiot.
\newblock Influence of network topology on protocol simulation.
\newblock In {\em International Conference on Networking}, pages 762--770.
  Springer, 2001.

\bibitem{Medina2001}
Alberto Medina, Anukool Lakhina, Ibrahim Matta, and John Byers.
\newblock {BRITE: Universal Topology Generation from a User's Perspective}.
\newblock Technical report, Boston, MA, USA, 2001.

\bibitem{Milic2009}
Bratislav Milic and Miroslaw Malek.
\newblock {NPART-node placement algorithm for realistic topologies in wireless
  multihop network simulation}.
\newblock In {\em Proceedings of the 2nd international conference on simulation
  tools and techniques}, 2009.

\bibitem{Nowak2014}
S~Nowak, M~Nowak, and K~Grochla.
\newblock {Properties of Advanced Metering Infrastructure Networks'
  Topologies}.
\newblock {\em Network Operations and Management Symposium (NOMS), 2014 IEEE},
  pages 1--6, 2014.

\bibitem{Rossi2013}
R~Rossi, S~Fahmy, and N~Talukder.
\newblock {A Multi-Level Approach for Evaluating Internet Topology Generators}.
\newblock {\em 2013 IFIP Networking Conference}, pages 9 pp.--9 pp., 2013.

\bibitem{Sanni2013}
M~L Sanni, A~A Hashim, F~Anwar, G~S~M Ahmed, and S~Ali.
\newblock {How to model wireless mesh networks topology}.
\newblock {\em IOP Conference Series: Materials Science and Engineering},
  53(1):012037, 2013.

\bibitem{stone1974cross}
Mervyn Stone.
\newblock Cross-validatory choice and assessment of statistical predictions.
\newblock {\em Journal of the royal statistical society. Series B
  (Methodological)}, pages 111--147, 1974.

\bibitem{Waxman1988}
B.~M. Waxman.
\newblock Routing of multipoint connections.
\newblock {\em IEEE Journal on Selected Areas in Communications},
  6(9):1617--1622, Dec 1988.

\bibitem{7510878}
Y.~Xu, J.~Liu, Y.~Shen, X.~Jiang, and T.~Taleb.
\newblock Security/qos-aware route selection in multi-hop wireless ad hoc
  networks.
\newblock In {\em 2016 IEEE International Conference on Communications (ICC)},
  pages 1--6, May 2016.

\bibitem{7925687}
Y.~Xu, J.~Liu, O.~Takahashi, N.~Shiratori, and X.~Jiang.
\newblock Soqr: Secure optimal qos routing in wireless ad hoc networks.
\newblock In {\em Proceedings of IEEE Wireless Communications and Networking
  Conference (WCNC)}, pages 1--6, 2017.

\end{thebibliography}

\endgroup

\end{document}